\documentclass[AMS,STIX1COL]{WileyNJD-v2}

\articletype{\textcolor[rgb]{0.00,0.00,1.00}{AUTHOR's ACCEPTED MANUSCRIPT}}%

\received{11 January 2021}

\begin{document}

\title{Stick-slip and convergence of feedback-controlled systems with Coulomb friction}

\author{Michael Ruderman*}



\authormark{MICHAEL RUDERMAN}

\address{\orgdiv{Faculty of Engineering and Science}, \orgname{University of Agder}, \orgaddress{\state{P.B. 422, Kristiansand, 4604}, \country{Norway}}}



\corres{*Corresponding address. \email{michael.ruderman@uia.no}}


\abstract[Summary]{An analysis of stick-slip behavior and
convergence of trajectories in the feedback-controlled motion
systems with discontinuous Coulomb friction is provided. A
closed-form parameter-dependent stiction region, around an
invariant equilibrium set, is proved to be always reachable and
globally attractive. It is shown that only asymptotic convergence
can be achieved, with at least one but mostly an infinite number
of consecutive stick-slip cycles, independent of the initial
conditions. Theoretical developments are supported by a number of
numerical results with dedicated convergence examples.}



\maketitle



\section{Introduction}
\label{sec:1}

Feedback-controlled motion systems are mostly subject to nonlinear
friction, and the direction-dependent Coulomb friction force plays
a crucial role owing to a (theoretical) discontinuity at velocity
zero-crossings. Although the more complex dynamic friction laws
(see, e.g., \cite{armstrong1994,Awrejcewicz2005,al2008} and
references therein) allow the frictional discontinuity to be
bypassed during analysis, the basic Coulomb friction phenomenon
continues to represent the same challenges in terms of a
controller convergence, especially in the presence of an integral
control action. An associated stick-slip behavior and so-called
frictional limit cycles were formerly addressed in
\cite{armstrong1996}. An algebraic prediction of stick-slip, with
a large set of parametric equalities, was compared to the
describing function method, while the Coulomb plus static friction
law was assumed for avoiding discontinuity at the velocity
zero-crossing. An explicit solution for friction-generated limit
cycles has also been proposed in \cite{olsson2001}, necessitating
static friction approximation (to avoid discontinuity) and also
requiring the stiction friction which is larger than the Coulomb
friction level. Despite including explicit analysis of state
trajectories for both sticking and slipping phases, no
straightforward conclusions about the appearance and convergence
of stick-slip behavior have been reported. Also several studies on
adaptive friction control, correspondingly estimation, attempted
formerly to address the nonlinear effects of friction and,
correspondingly, compensate for them, see e.g. \cite{canudas1987}.
The appearance of friction-induced (so-called hunting) limit
cycles has been briefly addressed in \cite{hensen2003}, for the
assumed LuGre \cite{DeWit1995} and so-called switch
\cite{Karnopp1985} friction models. Note that before, an earlier
analysis of stick-slip behavior and associated friction-induced
limit cycles can be found in \cite{radcliffe1990}. An explanation
of how a proportional-feedback-controlled motion with Coulomb
friction comes to sticking was subsequently shown in
\cite{alvarez2000} by using the invariance principle. Stick-slip
behavior, as an observable phenomenon known in the control
practice, was highlighted already in \cite{armstrong1994}, and
several following control studies have since there attempted to
analyze and compensate such behaviors. For instance, a related
analysis of under-compensation and over-compensation of the static
friction was reported in \cite{putra2007}. Issues associated with
a slow (creeping-like) convergence of the feedback-controlled
motion in presence of the Coulomb friction have been addressed and
experimentally demonstrated in \cite{ruderman2016}. More recently,
the convergence problems of a PID feedback control have been well
demonstrated with an accurate experiment in \cite{beerens2019},
while attempting to reduce the settling errors by a reset integral
control \cite{Clegg1958}. The related analysis has been also
reported before in \cite{bisoffi2017}. Despite a number of
experimental observations and elaborated studies reported in the
literature, it appears that no yet general consensus has been
established in relation to the friction-induced stick-slip cycles
in the feedback-controlled systems with Coulomb friction. In
particular, questions arise over when and under which conditions
the stick-slip cycles occur, and how a PID-controlled motion will
converge to zero equilibrium in the presence of Coulomb friction,
especially with discontinuity. Note that the problem of a slow
convergence in vicinity to a set reference position is of
particular relevance for the advanced motion control systems, see,
e.g., \cite{ruderman2020}. Yet, in the PID design and tuning, see,
e.g., \cite{ang2005}, the associated issues are not widely
accepted and have still to be formalized, this despite a huge
demand coming from a precision control engineering. This gap,
however, should not come as a fully surprising, given the fact of
a nontrivial friction microdynamics (visible from several
experimental studies \cite{koizumi1984,symens2005,yoon2019}), and
the uncertain and time-varying friction behavior, see e.g.
\cite{ruderman2015}.

Despite the appearance of the few papers mentioned above, a
clearly comprehensible analysis and explanation of the stick-slip
behavior due to the integral feedback effect in the presence of
Coulomb friction remains underexposed in the system and control
literature. The main objective of this paper is in filling this
gap. The work is dedicated to the contribution to the convergence
analysis of the feedback-controlled systems in the presence of the
Coulomb friction and, thus, to the understanding of stick-slip
cycles that occur in servomechanisms. The main contributions can
be highlighted as following: (i) we derive and describe the
closed-form parameter-dependent stiction region encompassing
equilibrium region, (ii) we prove that only asymptotic convergence
to this region can be achieved and that with stick-slip
oscillations. In order to keep the analysis general as possible
and to clarify the principal phenomenon of frictional-driven
stick-slip response, a classical Coulomb friction law with
discontinuity is assumed. This (unavoidably) led to a
variable-structure system dynamics, distinguishing between the
modes of a motion sticking and slipping. At the same time, we show
that all state trajectories always remain continuous and almost
always differentiable (except finite switching between both
modes). We provide theorems and identify the conditions to
demonstrate the sticking region around zero equilibrium to be
reachable and globally attractive. The developed analysis is
further reinforced by several illustrative numerical examples.

\subsection{Problem statement}
\label{sec:1:sub:1}

Throughout the paper, we will deal with the feedback-controlled
systems described by
\begin{equation}\label{eq:1}
   \ddot{\phi}(t) + K_{d} \dot{\phi}(t) + K_{p} \phi(t) + K_{i} \int \phi(t) dt + F(t)=0,
\end{equation}
where the derivative, proportional and integral feedback gains are
$K_{d}$, $K_{p}$ and $K_{i}$, respectively. Note that we are
purposefully focusing on a PID-type feedback control \eqref{eq:1},
where the integral control action is particularly critical for the
friction-driven stick-slip effects, as since long known in the
control practice. Other types of the feedback controls, still
including an integral control action, are also thinkable for
analysis but would go far beyond the provided analysis and
results. The nonlinear friction (that with discontinuity) is
denoted by $F$, and the set-point control problem is reduced to
the convergence problem for a non-zero initial condition, i.e.,
$\phi(0) \neq 0$. Furthermore, we use the following
simplifications of the system plant without loss of generality:
the relative motion of an inertial body with unity mass is
considered in the generalized $(\phi,\dot{\phi})$ coordinates. The
inherent system damping (including linear viscous friction) and
stiffness (of restoring spring elements) are incorporated (if
applicable) into $K_{d} > 0$ and $K_{p} > 0$, respectively. There
are no actuator constraints, so that the feedback of integral
output error is directly applicable via the gain factor $K_{i}>0$.

The control problem \eqref{eq:1} has long been associated with
issues of a slow and/or cyclic convergence of $\phi(t)$ in the
vicinity of steady-state for the set-point reference. This
(sometimes called hunting behavior or even hunting limit cycles)
has been addressed in analysis and also observed in several
controlled positioning experiments, e.g.,
\cite{armstrong1994,armstrong1996,olsson2001,hensen2003,ruderman2016,beerens2019}.
The phenomena seem to be associated with an integral control
action and nonlinear (Coulomb-type) friction within a vanishing
region around the equilibria, where the potential field of
proportional feedback weakens and cannot provide $\phi(t)
\rightarrow 0$ within the certain application-required time $t <
\mathrm{const}$. The hunting behavior is directly cognate with
stick-slip, where a smooth (continuous) motion alternates with a
sticking phase of zero or slowly creeping displacement. Stick-slip
appearance, parametric conditions, and convergence in semi-stable
limit cycles are the focus of our study, while we assume the
Coulomb friction force with discontinuity.

\section{Stiction due to discontinuous Coulomb friction}
\label{sec:2}

In this Section, we analyze the stick-slip behavior of the
\eqref{eq:1} system, for which the classical Coulomb friction with
discontinuity is represented by $F(\dot{\phi}) = F_{c} \,
\mathrm{sign}(\dot{\phi})$. Here, the Coulomb friction coefficient
is $F_{c} > 0$, and the sign operator is defined by
\begin{equation}\label{eq:2}
    \mathrm{sign}(z)= \left\{%
\begin{array}{ll}
    1, & \; z>0, \\
    \left[1, 1\right], & \; z=0,\\
    -1, & \; z<0. \\
\end{array}%
\right.
\end{equation}
Note that \eqref{eq:2} constitutes an ideal relay with
instantaneous switching upon change of the input sign. We also
note that for a zero-displacement rate, the friction equation
becomes an inclusion $F(0) \in [-F_{c}, F_{c}]$ in the Filippov
sense \cite{filippov1988}, when one is seeking for the
corresponding analytic solution.

We will consider the feedback-controlled system in a minimal
state-space representation as follows:
\begin{eqnarray}
\label{eq:3}
  \dot{x} &=& Ax+Bu, \\
\label{eq:4}
   y      &=& C x,\\
\label{eq:5}
   u      &=&-\mathrm{sign}(y).
\end{eqnarray}
Note that in this way, we also approach the system notation
provided in \cite{johansson1999} for analysis of the relay
feedback systems (RFSs). Introducing the state vector $x=(x_{1},
x_{2}, x_{3})^{T}\in \mathbb{R}^{3}$ of the integral, output, and
derivative errors, \eqref{eq:1} can be rewritten as
\eqref{eq:3}-\eqref{eq:5}, with the system matrix
\begin{equation}\label{eq:6}
   A= \left(%
\begin{array}{ccc}
  0 & 1 & 0 \\
  0 & 0 & 1 \\
  -K_{i} & -K_{p} & -K_{d} \\
\end{array}%
\right),
\end{equation}
and input and output distribution vectors
\begin{equation}\label{eq:7}
B=\left(%
\begin{array}{c}
  0 \\
  0 \\
  F_{c} \\
\end{array}%
\right), \quad C^{T}=\left(%
\begin{array}{c}
  0 \\
  0 \\
  1 \\
\end{array}%
\right)
\end{equation}
correspondingly.

\subsection{Without integral feedback}
\label{sec:2:sub:1}

\begin{figure}[!h]
\centering
\includegraphics[width=0.6\columnwidth]{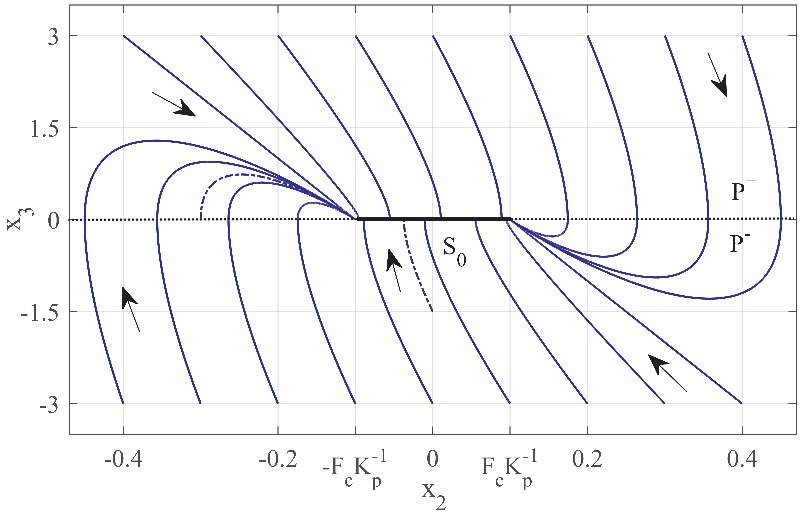}
\caption{Phase portrait of $(x_{2}, x_{3})$-trajectories of the
\eqref{eq:3}-\eqref{eq:7} system without integral feedback,
attracted to $S_0$ from various initial values} \label{fig:1}
\end{figure}

Firstly, we consider the system \eqref{eq:3}-\eqref{eq:7} without
an integral feedback action, meaning $K_{i}=0$. In this case, the
phase-plane $(x_{2}, x_{3}) \in \mathbb{R}^{2}$ is divided into
two regions
\begin{equation}\label{eq:8}
    P^{+}=\{x\in\mathbb{ R}^{2} : x_{3}> 0\}, \quad P^{-}=\{x\in
    \mathbb{R}^{2} : x_{3}<0\}
\end{equation}
by the discontinuity manifold $S=\{x\in \mathbb{R} ^{2}
:x_{3}=0\}$. It can be seen that in the discontinuity manifold
$S$, the vector fields of the state value $x_s$ \footnote{Note
that in the following we will often use: (i) the subscript or
superscript character $s$ for denoting the sticking phase and
correspondingly the sliding mode, and (ii) the subscript or
superscript character $c$ for denoting the slipping phase and
correspondingly continuous mode. Both will be used for the time
argument $t$ and the state variables $x$, correspondingly
$x_1,\,x_2,\, x_3$.} are given by
\begin{eqnarray}
\label{eq:9}
      f^{+} (x_{s}) & = \overset{x \in P^{+}} {\underset{x \rightarrow x_s} \lim}  (Ax+Bu)=\left(%
\begin{array}{c}
  0 \\
  -K_{p} x_{2} -F_{c}
\end{array}%
\right), \\
\label{eq:10}
      f^{-} (x_{s}) & = \overset{x \in P^{-}} {\underset{x \rightarrow x_s} \lim} (Ax+Bu)=\left(%
\begin{array}{c}
  0 \\
  -K_{p} x_{2} +F_{c}
\end{array}%
\right),
\end{eqnarray}
and are pointing in the opposite directions within $|x_{2}| \leq
F_{c} K_{p}^{-1}$. On the contrary, outside of this region
(denoted by $S_{0}$ in Figure \ref{fig:1}), both vector fields are
pointing in the same direction, towards $P^{+}$ for $x_{2}< -F_{c}
K_{p}^{-1}$ and towards $P^{-}$ for $x_{2} > F_{c}K_{p}^{-1}$.
Since both vector fields are normal to the manifold $S$, neither
smooth motion nor sliding mode can occur for the $(x_{2},
x_{3})\in S_{0}$ trajectories. It means that any trajectory
reaching $S_{0}$ will remain there $\forall \: t \rightarrow
\infty$. Therefore, $S_{0}$ constitutes the largest invariant set
of equilibrium points, for \eqref{eq:3}-\eqref{eq:7} without
integral control action. Note that this has also been shown in
\cite{alvarez2000} and is well known when a relative motion with
Coulomb friction is controlled by the proportional-derivative (PD)
feedback only. In this case, the set value error can be reduced by
increasing $K_{p}$ but cannot be driven to zero as long as $F_{c}
\neq 0$. The phase portraits of the trajectories converging to
$S_{0}$ are exemplary shown in Figure \ref{fig:1} and marked with
arrows.

\subsection{With integral feedback}
\label{sec:2:sub:2}

When allowing for $K_{i} \neq 0$, it is intuitively apparent that
having reached a point $x(t_{s}) \in S_{0}$ at $t = t_{s}$, the
trajectory cannot remain there for all times $t_{s} < t < \infty$.
While the motion states $x_{3}(t_{s})=0$ and $x_{2}(t_{s}) =
\mathrm{const} \neq 0$, the integral control effort $K_{i}
x_2(t_{s}) \int^{t}_{t_{s}} dt$ grows continuously and, at some
finite time ($t_{c} > t_{s}$), will lead to the breakaway
\cite{ruderman2017} and a new onset of a continuous motion. This
alternating phase, upon the system sticking, is often referred to
as \emph{slipping}, cf., e.g., \cite{olsson2001}, so that a
stick-slip motion \cite{armstrong1994,al2008} appears also in form
of the limit cycles. In order to analyze the friction-induced
limit cycles, sometimes denoted as hunting-limit cycles, cf.
\cite{hensen2003}, we firstly need to look into the system
dynamics during the system stiction, i.e., for $t_{s} < t <
t_{c}$. Here we recall that during a stiction phase, the system
\eqref{eq:3}-\eqref{eq:7} produces a continuous switching (with
infinite frequency) once $x_{3} = 0$, this owing to the
discontinuous relay nonlinearity \eqref{eq:2} which is acting in
the feedback loop. One can also notice, in upfront, that for
$x_{3} = 0$ the solution of \eqref{eq:3}-\eqref{eq:7} is needed to
be specified in the Filippov sense \cite{filippov1988}. Further we
note that the below given developments are motivated by analysis
of existence of the fast switches provided in \cite{johansson1999}
for RFS, while the obtained original results rely on the
sliding-mode principles, see, e.g.,
\cite{edwards1998,shtessel2014}.

Consider the switching variable (or more generally surface) $S =
Cx(t) = 0$, for which the sliding mode should occur on the
manifold $S$. This requires that the existence and reachability
condition, cf. \cite{edwards1998},
\begin{equation}\label{eq:11}
    \dot{S} S \leq - \eta |S|
\end{equation}
is fulfilled, where $\eta$ is a small positive constant.

\begin{theorem} 
\label{thm:1} Given is the control system
\eqref{eq:3}-\eqref{eq:7} with the Coulomb friction. The system is
sticking at $x_{3}=0$ iff
\begin{equation}\label{eq:12}
    |K_{i}x_{1}| + |K_{p}x_{2}| \leq F_{c}.
\end{equation}

\end{theorem}

\begin{proof}
The system remains sticking as long as it is in the sliding mode
for which \eqref{eq:11} is fulfilled. The sliding-mode condition
\eqref{eq:11} can be rewritten as
\begin{equation}\label{eq:13}
    \dot{S} \, \mathrm{sign} (S) \leq -\eta,
\end{equation}
while the time derivative of the sliding surface is
\begin{equation}\label{eq:14}
    \dot{S} = (C A x \pm CB) = \bigl(C A x - \mathrm{sign}(S) CB \bigr),
\end{equation}
depending on the sign of $Cx$. Substituting \eqref{eq:14} into
\eqref{eq:13} results in
\begin{eqnarray}
\label{eq:15}
  CA x  & \leq & CB-\eta \qquad \hbox{for} \quad \mathrm{sign}(S)>0, \\
  -CA x & \leq & CB-\eta \qquad \hbox{for} \quad \mathrm{sign}(S)<0.
\label{eq:16}
\end{eqnarray}
Since $CB, \eta > 0$, the inequalities \eqref{eq:15} and
\eqref{eq:16} can be summarized in
\begin{equation}\label{eq:17}
    |CAx| \leq CB -\eta.
\end{equation}
Evaluating \eqref{eq:17} with $x_{3}=0$ and $0 \neq \eta
\rightarrow 0^{+}$ results in \eqref{eq:12} and completes the
proof.
\end{proof}

\begin{remark}
The condition obtained by the Theorem \ref{thm:1} is equivalent to
the set of attraction $\{x\in S:|CAx|<|CB|\}$ for $CB>0$ that was
demonstrated in \cite[Section~4]{johansson1999}.
\end{remark}

Now, we are interested in the state dynamics during the system
stiction, which means within the sliding mode. Since staying in
the sliding mode (correspondingly on the switching surface $S
\equiv 0$) requires
\begin{equation}\label{eq:18}
    \dot{S}=C\dot{x}=CA x+CB u=0 \qquad \hbox{for} \quad t_{s}<t<t_{c},
\end{equation}
one obtains the so-called equivalent control as
\begin{equation}\label{eq:19}
    u_{e}=-(CB)^{-1} CAx.
\end{equation}
Recall that an equivalent control, \cite{shtessel2014}, is the
linear one (i.e. without a relay action) which is required to
maintain the system in an ideal sliding mode without
fast-switching. Consequently, substituting \eqref{eq:19} into
\eqref{eq:3} results in the equivalent system dynamics
\begin{equation}\label{eq:20}
    \dot{x}_e=\bigl[I-B(CB)^{-1}C \bigr] A x_e = O A x_e,
\end{equation}
which governs the state trajectories as long as the system remains
in the sliding mode, and where $x_e = (x_1, x_2, 0)^T$. Here $O$
is the so-called projection operator of the original system
dynamics, satisfying the properties $C O = 0$ and $ O B = 0$.
Evaluating \eqref{eq:20} with \eqref{eq:6} and \eqref{eq:7} yields
the equivalent system dynamics during the stiction as
\begin{equation}\label{eq:21}
    \left(%
\begin{array}{c}
  \dot{x}_{1} \\
  \dot{x}_{2} \\
  \dot{x}_{3}\\
\end{array}%
\right) =
\left(%
\begin{array}{ccc}
  0 & 1 & 0 \\
  0 & 0 & 1 \\
  0 & 0 & 0 \\
\end{array}%
\right)
\left(%
\begin{array}{c}
  x_{1}(t_{s}) \\
  x_{2}(t_{s}) \\
  0 \\
\end{array}%
\right).
\end{equation}
It can be seen that neither relative displacement nor its rate
will change when the system is sticking, although the integral
error grows according to
\begin{equation}\label{eq:22}
    x_{1}(t) = x_{1}(t_{s}) + x_{2}(t_{s}) \int \limits^{t_c }_{t_s}dt.
\end{equation}
Further it can be noted that if $K_{i}=0$ then the condition
\eqref{eq:12}, correspondingly the inequality $|CAx| \leq |CB|$,
reduces to $|x_{2}| \leq F_{c}K_{p}^{-1}$, while the sliding mode
\eqref{eq:21} reduces to the zero dynamics of the system in
stiction (cf. with results in Section \ref{sec:2:sub:1}).

\subsection{Region of attraction}
\label{sec:2:sub:3}

Theorem \ref{thm:1} provides the necessary and sufficient
condition for the system \eqref{eq:3}-\eqref{eq:7} remains
sticking. Yet it is also necessary to demonstrate the global
attraction of state trajectories to the stiction region. Recall
that the latter corresponds to the subset
\begin{equation}\label{eq:23}
    S_{0} = \{x \in \mathbb{R}^{3}\, : \, x_{3}=0,\, |K_{i} x_{1}| + |K_{p} x_{2} | \leq F_{c} \}
\end{equation}
where the sliding mode occurs (cf. \eqref{eq:12} and
\eqref{eq:21}).

\begin{figure}[!h]
\centering
\includegraphics[width=0.5\columnwidth]{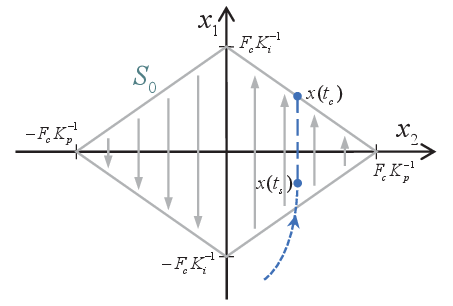}
\caption{Rhombus-shape, in $(x_1,x_2)$-projection, of the
$S_0$-region of attraction, with vector-field during the stiction
mode and example of an entering and leaving trajectory at $t_s$
and $t_c$, respectively} \label{fig:2}
\end{figure}

Firstly, we will explore the persistence of the sliding mode,
meaning we will prove whether the system can stay incessantly
inside of $S_0^s = \{ S_0 \, : \, x_2 \neq 0 \}$, i.e., for all
times $t_s < t \rightarrow \infty$. By making $(x_{1},
x_{2})$-projection of $ x \in \mathbb{R}^{3}$, one can show that
\eqref{eq:12} results in a rhombus, as schematically illustrated
in Figure \ref{fig:2}.
The indicated vector field is unambiguous due to the integral
control action (cf. the sliding-mode dynamics \eqref{eq:21}). It
means that after reaching $S_{0}^s$ at $t_{s}$, any trajectory
will leave it at $t_{c}$ once it hits the boundary of $S_{0}$.
Denoting the point of reaching $S_{0}^s$ by $x(t_{s}) \equiv
(x^{s}_{1}, x^{s}_{2}, 0)$, one can calculate the new point of
leaving $S_{0}^s$ as
\begin{equation}\label{eq:24}
    x_{1}(t_{c}) = : x^{c}_{1} = x^{s}_{2} \bigl( F_{c} |x^{s}_{2}|^{-1} -K_{p} \bigr) K^{-1}_{i}.
\end{equation}
Correspondingly, from \eqref{eq:22} and \eqref{eq:24} one obtains
the time of leaving $S_{0}^s$ as
\begin{equation}\label{eq:25}
    t_{c} = \Bigl[ x^{s}_{2} \bigl( F_{c}|x^{s}_{2}|^{-1}-K_{p} \bigr) K^{-1}_{i} - x^{s}_{1} + x^{s}_{2} t_{s} \Bigr] (x^{s}_{2})^{-1}.
\end{equation}
From the above, it can be recognized that if $K_{i} \rightarrow
0$, the stiction region $S_{0}$ blows to the entire $(x_{1},
x_{2})$-subspace and, consequently, $t_{c} \rightarrow \infty$. It
means that a system trajectory will never leave $S_{0}$ having
reached it -- the result which is fully in line with what was
demonstrated in Section \ref{sec:2:sub:1}. On other hand, if
allowing for $K_{i} \rightarrow \infty$ the time instant $t_{c}
\rightarrow t^{+}_{s}$, according to \eqref{eq:25} with $x^{s}_{1}
\rightarrow 0$, due to $S_{0}$ is collapsing to
$\mathrm{proj}_{x_{2}} S_{0}$.

Let us now demonstrate that $S_{0}$ is globally attractive for all
initial values outside of $S_{0}$, meaning $\forall \: x(t_{0})\in
\mathbb{R}^{3} \backslash S_{0}$. Using the eigen-dynamics of
\eqref{eq:3} with \eqref{eq:6}, which are linear, one can ensure
the global exponential stability by analyzing the characteristic
polynomial
\begin{equation}\label{eq:26}
    s^{3}+K_{d}s^{2}+K_{p}s + K_{i} = 0
\end{equation}
and applying the standard Routh-Hurwitz stability criterion. Then,
the control parameters condition
\begin{equation}\label{eq:27}
    K_{d} K_{p} > K_{i}
\end{equation}
should then be satisfied, for guaranteeing that all eigenvalues
$\lambda_{i}$ of the system matrix \eqref{eq:6} have $\mathrm{Re}
\{ \lambda_{i} \} < 0$ with $i=1,\ldots,3$. Then, the resulting
(switched) subsystems $\dot{x} = Ax \mp B$ behave as
asymptotically stable in both subspaces $\{x\in \mathbb{R}^{3}
\backslash S_{0}:x_{3}\gtrless 0\}$ correspondingly. It should be
noted that the condition of the above parameters is conservative,
since the Coulomb friction itself is always dissipative,
independently of whether $x_{3} > 0$ or $x_{3} < 0$. This can be
shown by considering the dissipated energy
\begin{equation}\label{eq:28}
    \bar{V}(t) = -F(t) \int \dot{\phi}(t) dt = - F(t)\bar{\phi},
\end{equation}
which is equivalent to a mechanical work provided by the constant
friction force $F$ along an unidirectional displacement
$\bar{\phi}$. Taking the time derivative of \eqref{eq:28} and
substituting the Coulomb friction law results in
\begin{eqnarray}
\nonumber  \frac{d}{d t} \bar{V}(t) &=& - \frac{d}{d t} F(t) \bar{\phi}  - F(t) \frac{d}{d t} \bar{\phi} \\[0.2mm]
   &=& 0 - F_{c} \mathrm{sign} \bigl( \dot{\phi}(t) \bigr) \dot{\phi}(t) = -F_{c} \bigl|\dot{\phi}(t)\bigr|.
\label{eq:29}
\end{eqnarray}
Therefore, $\dot{\bar{V}}(t) < 0$ for all $x_{3}(t) \neq 0$. This
quite intuitive, yet relevant to be analytically expressed,
condition reveals the relay feedback \eqref{eq:5} as an additional
(rate-independent) damping, which contributes to stabilization of
the closed-loop dynamics \eqref{eq:3}-\eqref{eq:7}. This result
will be further used for the proof of Corollary \ref{clr:1}.
Notwithstanding this additional stabilizing by-effect, we will
keep the conservative stability condition \eqref{eq:27} as the
sufficient (but not necessary) one. This appears reasonable due to
an usually uncertain Coulomb friction coefficient and, hence, in
order for increasing the overall robustness of the feedback
control system. The following example should, however, exemplify
the additionally stabilizing behavior of the Coulomb friction,
even when \eqref{eq:27} is violated.

\begin{eexample}
\label{exmp:1}

Consider the system \eqref{eq:3}-\eqref{eq:7} with $K_{d}=0$,
$K_{p}=100$ and $K_{i}=1$. The eigenvalues of the system matrix
$A$ are $\lambda_{1}=-0.01$, $\lambda_{2,3} = 0.0005 \pm 10 j$,
which implies the linear subsystem is asymptotically unstable. It
should also be noted that \eqref{eq:27} is not fulfilled. To
evaluate the trajectories of the system \eqref{eq:3}-\eqref{eq:7},
one can use the particular solution
\begin{equation}\label{eq:30}
    x(t) = e^{At} x(\tau) + A^{-1}\bigl(e^{At}-I\bigr)Bu,
\end{equation}
for the constant control $u = \mp 1$, which corresponds to the
relay \eqref{eq:5}, switched in the $x_{3} > 0$ and $x_{3} < 0$
subspaces. The initial values $x(\tau) = [x_{1}(t), x_{2}(t),
0]^{T}$ at $t=\tau$ should be reassigned each time the trajectory
crosses the $(x_{1}, x_{2})$-plane, meaning the relay switches at
$x_{3} = 0$ outside of $S_{0}$. The $x_3$-state trajectory, with
an initial value $x_3(0)=10$, is shown in Figure \ref{fig:3}, once
for $F_{c}=0$ (solid red line) and once for $F_{c}=1$ (blue
dash-dot line).
\begin{figure}[!h]
\centering
\includegraphics[width=0.6\columnwidth]{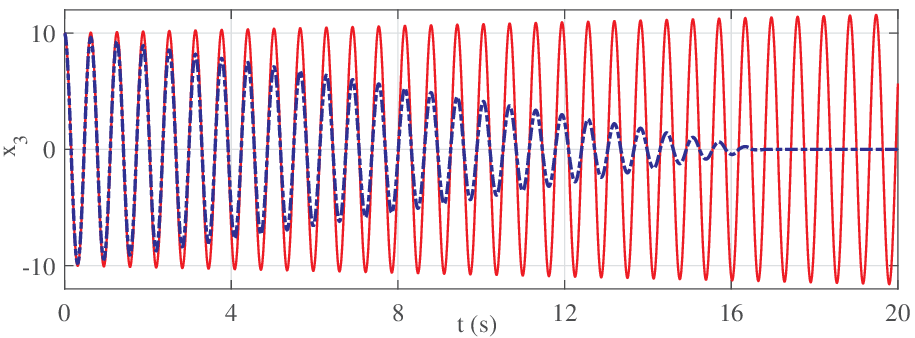}
\caption{Velocity trajectories (from $x_3(0)=10$) of the Example
\ref{exmp:1} system with $F_{c}=0$ (solid red line) and $F_{c}=1$
(blue dash-dot line)} \label{fig:3}
\end{figure}
It is easy to recognize that even a low-valued Coulomb friction
coefficient ($F_{c}=1$ compared to the proportional feedback gain
$K_{p}=100$) leads to a stabilization of the, otherwise, unstable
closed-loop control response.
\end{eexample}

\begin{corollary}
\label{clr:1}

Consider the system \eqref{eq:3}-\eqref{eq:7} with the control
parameters satisfying \eqref{eq:27}. The stiction region
\eqref{eq:23}, given by the Theorem \ref{thm:1}, is globally
attractive for all initial values outside of this region, i.e. for
all $x(t_{0}) \in \mathbb{R}^{3} \backslash S_{0}$.

\end{corollary}

\begin{proof}

By virtue of the passivity theorem, e.g.,
\cite{zames1966,khalil2002}, the feedback interconnection of the
energy dissipating systems is also energy dissipating. Since
\eqref{eq:3}, \eqref{eq:4} is dissipative when \eqref{eq:27} is
fulfilled, and \eqref{eq:5} is also dissipative for $x_{3} \neq
0$, their feedback interconnection yields dissipative almost
everywhere (except $x_{3} = 0$) outside of $S_{0}$. This implies
that any $x(t)$-trajectory, starting from outside of $S_{0}$,
converges continuously, and some ball $\mathcal{B} \equiv \|x\|$
around the origin shrinks over time:
\begin{equation}\label{eq:31}
    \|x(t_{2})\| < \|x(t_{1})\| \quad \forall \quad t_{2} > t_{1}, \:
    x  \in \mathbb{R}^{3} \backslash S_{0}.
\end{equation}
For some $t_{3}>t_{2}$, the shrinking circle becomes
$\mathrm{proj}_{(x_{1},x_{2})} \mathcal{B} \subseteq S_{0}$, and
for $t_{4} \geq t_{3}$ a zero velocity $x_{3}(t_{4}) = 0$ will be
consequently reached. This implies $x(t_{4}) \in S_{0}$, which
completes the proof.

\end{proof}

\begin{remark}
The sliding-mode condition \eqref{eq:13}, which results in $|CAx|
\leq CB$ and proves the Theorem \ref{thm:1}, correspondingly,
constitutes the existence and reachability condition for $S_{0}$,
and is necessary but not sufficient. This is because \eqref{eq:12}
does not contain any requirements imposed on the $K_{d}$-parameter
value. Theorem \ref{thm:1} and Corollary \ref{clr:1} constitute
the necessary and sufficient conditions for $S_{0}$ to be both --
the globally reachable and attractive from outside of $S_{0}$.
\end{remark}

\section{Analysis of stick-slip convergence}
\label{sec:3}

\begin{figure}[!h]
(a) \hspace{0.1\columnwidth}
\includegraphics[width=0.5\columnwidth]{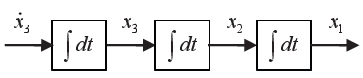}
\vspace{0.2cm} \\
(b)
\includegraphics[width=0.25\columnwidth]{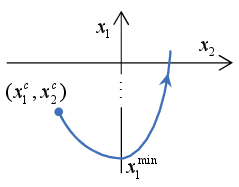}
(c)
\includegraphics[width=0.25\columnwidth]{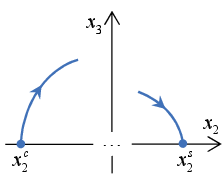}
\vspace{0.1cm} \\
(d) \hspace{0.07\columnwidth}
\includegraphics[width=0.5\columnwidth]{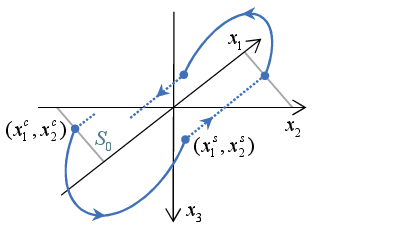} \\
\centering \caption{Phase portrait at stick-slip; (a)
triple-integrator chain, (b) $(x_1,x_2)$-projection during
sticking, (c) $(x_2,x_3)$-projection during slipping, (d) typical
trajectory during one stick-slip cycle} \label{fig:4}
\end{figure}

In this Section, we will analyze the convergence behavior of
stick-slip trajectories of the system \eqref{eq:3}-\eqref{eq:7}.
Recall that having reached $S_{0}^{s}$ at $t_{s}$, the $x(t)$
trajectory will leave it at $t_{c}$, given by \eqref{eq:25}, which
is due to the growing $|x_{1}(t)|$ value, that will (unavoidably)
violate the stiction condition \eqref{eq:12}. To show
(qualitatively) how the state trajectories evolve during a
stick-slip cycle, consider the triple-integrator chain (see Figure
\ref{fig:4}(a)), which arises out of the closed-loop dynamics
\eqref{eq:1}.
Eliminating the time argument, which is a standard procedure for a
phase-plane construction, one can write
\begin{equation}\label{eq:32}
    x_{n}dx_{n}=\dot{x}_{n}dx_{n-1},\quad n = \{3, 2\},
\end{equation}
in general terms, that for the first and second (from the left to
the right) integrator. For an unidirectional motion (here
$\mathrm{sign}(x_{3})=1$, for instance, is assumed) and piecewise
constant approximation $\dot{x}_{n}=\mathrm{const}$, one obtains
\begin{eqnarray}
\label{eq:33}
  x_{1} &=& x^{2}_{2}(2x_{3})^{-1}-x_{1}^{min} ,\\
  x_{2} &=& x^{2}_{3}(2\dot{x}_{3})^{-1}-x_{2}^{c}
\label{eq:34}
\end{eqnarray}
after integrating the left- and right-hand sides of \eqref{eq:32}.
Obviously, for $t\geq t_{c}$, the $x_{1}(t)$-trajectory evolves
parabolically, depending from $x_{2}(t)$, see Figures
\ref{fig:4}(b). The $x_{3}(t)$-trajectory is square-root-dependent
of $x_{2}(t)$, see Figures \ref{fig:4}(c) correspondingly. Note
that the increasing and decreasing segments of the corresponding
trajectories are both asymmetric, effectively due to non-constant
$\dot{x}_{n}$-value as the motion evolves. At the same time, one
can stress that the extremum $x_{1}^{min}$ (here, minimal due to
the assumed positive sign of velocity) always lies on the
$x_{1}$-axis (cf. Figure \ref{fig:4}(b)) owing to $x_{1}=\int
x_{2}dt$, and $\mathrm{sign}(x_3)=\mathrm{const}$. Differently,
the $(x_{2},x_{3})$-projection of the $x(t)$-trajectory can be
shifted along the $x_{2}$-axis, while it always ends in
$x_{3}(t_{s})=0$ for $x(t)\in S_{0}$ (cf. Figure \ref{fig:4}(c)).
The resulting alternation of the stick-slip phases is
schematically shown in Figure \ref{fig:4}(d).

\begin{proposition}

Having reached $S_{0}$, the system \eqref{eq:3}-\eqref{eq:7} does
not leave $\Omega \in \mathbb{R}^{3}$ with
$\mathrm{proj}_{(x_{1},x_{2})} \Omega \subseteq S_{0}$ and
converges \emph{asymptotically} to $x(t) \underset{t \rightarrow
\infty}{=} \bigl \{ (x_1, 0, 0): \, |x_{1}| \leq F_{c}K_{i}^{-1}
\bigr\} $ within multiple (or at least one) stick-slip cycles. The
stick-slip cycles can occur with either zero-crossing of $x_{2}$
or with keeping the same $\mathrm{sign}(x_{2}(t_{s,1}))$, where
$t_{s,1}$ is the time instant when $x(t)$ reaches $S_{0}$ for the
first time.

\end{proposition}

When first disregarding the frictional side-effect, i.e.,
$F_{c}=0$, it is well understood that a non-overshoot of the set
reference value cannot be reached, independently of the assigned
control parameters, provided (i) $K_{i} \neq 0$ and (ii) the
initial conditions are such that either $x_{1}(0)=0$ or
$\mathrm{sign}(x_{1}(0)) = \mathrm{sign}(x_{2}(0))$. This becomes
evident since the integral error state $x_{1}(t)$ accumulates the
output error over time. That is, in order for the $|x_{1}(t)|$
starts to decrease, at least one change of the $x_{2}(t)$-sign is
required. An exception is when $\mathrm{sign}(x_{1}(0)) \neq
\mathrm{sign}(x_{2}(0))$, which allows both $x_{1}(t)$ and
$x_{2}(t)$ to converge to zero from the opposite directions. Thus,
at least one overshoot should appear, even if all control gains
are assigned to have the real poles only; see later the Example
\ref{exmp:4}.

When the Coulomb friction becomes effective, i.e., $F_{c} \neq 0$,
the system can change to the stiction again, and that also without
overshoot of $x_2=0$, after starting to slip at $x(t_{c})$. It
means that a motion trajectory lands again onto $S_{0}$ at time
$t_{s,2} > t_{c} > t_{s,1}$ and that with
$\mathrm{sign}(x_{2}(t_{s,2})) = \mathrm{sign}(x_{2}(t_{s,1}))$.
Since the system with $K_{d},F_{c}
> 0$ is dissipative, the energy level is $V(t_{s,2}) < V(t_{c})$,
meaning the motion trajectory $x(t)$ always lands onto $S_{0}$
closer to the origin than it was when leaving $S_{0}$ in
$x(t_{c})$. Note that the system energy within $S_{0}$ can be
expressed by the potential field of the proportional and integral
control errors, yielding
\begin{equation}\label{eq:35}
    V(t)=\frac{1}{2}K_{i}x_{1}^{2}+\frac{1}{2}K_{p}x_{2}^{2} \quad
    \hbox{for} \quad t \in [t_{s},\ldots,t_{c}].
\end{equation}
One can recognize that the energy level \eqref{eq:35}, of the
system in stiction, is an ellipse
\begin{equation}\label{eq:36}
    \frac{x_{2}^{2}}{a^{2}}+\frac{x_{1}^{2}}{b^{2}}=1 \quad
    \hbox{with} \quad a^{2}= 2V K_{p}^{-1}, \;  b^{2}=2V K_{i}^{-1}.
\end{equation}
Since the energy $V(t_{c})$ is bounded by $S_{0}$, cf. Figure
\ref{fig:2}, one can show that the semi-major axis is $a\leq
F_{c}K_{p}^{-1}$ and the semi-minor axis is $b\leq
F_{c}K_{i}^{-1}$. From the system dissipativity and global
attractiveness of $S_{0}$, cf. Corollary \ref{clr:1}, it follows
that the trajectory becomes sticking again, meaning $x(t)\in
S_{0}$ for $t\geq t_{s,2}> t_{c}$. Since $V(t_{s,2}) < V(t_{c})$,
the ellipse \eqref{eq:36} shrinks as $a$ and $b$ become smaller;
note that both are proportional to $V(t)$. It is important to
notice that during the system is sticking, the energy level
increases, on the contrary, since $V(t_{c})>V(t_{s,1})$. This is
consequently logical since the integral control action feed an
additional energy into the control loop when the system is at a
standstill. That leads to a breakaway and allows for the motion to
restart again once the sticking trajectory reaches the
$S_{0}$-boundary.

\begin{remark}

When the state trajectory reattains the stiction region
$x(t_{s,2})\in S_{0}$ without overshoot, meaning
$\mathrm{sign}(x_{2}(t_{s,2})) = \mathrm{sign}(x_{2}(t_{c}))$, the
system is over-damped by the Coulomb friction. Otherwise,
$\mathrm{sign}(x_{2}(t_{s,2})) \neq \mathrm{sign} (x_{2}(t_{c}))$
means the system is said to be under-damped by the Coulomb
friction. A special, but as will be shown not feasible, case of
$x_{2}(t_{s,2})=0$, meaning the system reaches equilibrium $S_{0}
\backslash S_{0}^{s}$ and remains there $\forall \: t \geq
t_{s,2}$, is analyzed below by proving the Theorem \ref{thm:2}.

\end{remark}

\begin{theorem} 
\label{thm:2}

The system \eqref{eq:3}-\eqref{eq:7}, with control parameters
satisfying \eqref{eq:27} and $F_{c}>0$, converges asymptotically
to the invariant set $\Lambda =\{(x_{1},0,0):|x_{1}| \leq
F_{c}K_{i}^{-1}\}$ during a number of stick-slip cycles $N \in
\mathbb{N}$, with $1 \leq N < \infty$. And there are no system
parameter values and stick-slip initial conditions
$(x_{1}(t_{s,n}),x_{2}(t_{s,n}))$ with $n < N$ which allow the
trajectory to reach $\Lambda$ at the end of the next following
stick-slip cycle within the time $t_{s,n+1} > t_{c,n}> t_{s,n}$.

\end{theorem}

\begin{proof}

The convergence to $\Lambda$ follows from system the dissipativity
during the slipping and, correspondingly, shrinking ellipse
\eqref{eq:36}, which implies an always decreasing energy level by
the end of one stick-slip cycle, i.e. $V(t_{s,n+1})< V(t_{s,n})$.
This implies $|x_{2}(t_{s,n+1})|< |x_{2}(t_{s,n})|$ for $n\in N$
and ensures such $x(t)$-trajectories which start slipping at
$t_{c,n}$ and land closer to the origin at $t_{s,n+1}$ than before
at $t_{s,n}$.

The proof of the second part of the Theorem \ref{thm:2}, which
says it is impossible to reach the invariant equilibrium set
$\Lambda$ after one particular stick-slip cycle, follows through
the contradiction. For this purpose, we should first assume that
there is a particular setting $\bigl(K_{p}, K_{i}, K_{d}, F_{c},
x_{1}(t_{s,n}), x_{2}(t_{s,n})\bigr)$ for which the state
trajectory $x(t_{s,n+1}) \in \Lambda$, i.e. in the next stiction
phase at the finite time $t_{s,n+1}>t_{c,n}>t_{s,n}$. The initial
conditions of a slipping phase are always given, cf. with Section
\ref{sec:2:sub:3}, by
\begin{eqnarray}
\label{eq:37}
  x_{2}(t_{c,n}) &=& x_2(t_{s,n}), \\
\label{eq:38}
  x_{1}(t_{c,n}) &=& \frac{F_{c}}{K_{i}} - \frac{K_{p}}{K_{i}} x_{2}(t_{c,n}) \: \hbox{ in 1st quadrant,} \\
\label{eq:39}
  x_{1}(t_{c,n}) &=& - \frac{F_{c}}{K_{i}} - \frac{K_{p}}{K_{i}} x_{2}(t_{c,n}) \: \hbox{ in 3rd quadrant}
\end{eqnarray}
This becomes apparent when inspecting the stiction phase dynamics
\eqref{eq:21}, \eqref{eq:22} and $S_{0}$-boundary, cf. Figure
\ref{fig:2}. For reaching $\Lambda$ at a final time instant $\psi
= t_{s,n+1}$, while starting at $\tau = t_{c,n}$, an explicit
particular solution of
\begin{equation}\label{eq:40}
0 = C \Bigl [ e^{A \psi} \left(%
\begin{array}{c}
  x_{1}(\tau) \\[-0.05cm]
  x_{2}(\tau) \\[-0.05cm]
  0
\end{array}%
\right) + A^{-1}(e^{A \psi} -I ) B u - \left(%
\begin{array}{c}
  x_{1}(\psi) \\[-0.05cm]
  0 \\[-0.05cm]
  0
\end{array}%
\right) \Bigr]
\end{equation}
with $u= \pm 1$, should exist, cf. with \eqref{eq:30}. Due to the
symmetry of solutions, we will consider the 1st quadrant of $S_0$
only, i.e., with the above initial condition \eqref{eq:38} and $u
= + 1$ correspondingly, this without loss of generality when
solving \eqref{eq:40}. Recall that the matrix exponential
\begin{equation}\label{eq:41}
    e^{A \psi}=\sum^{\infty}_{k=0}\frac{(A \psi)^{k}}{k!}
\end{equation}
has to be evaluated to find an explicit solution of \eqref{eq:40}.
Substituting the initial conditions, i.e. \eqref{eq:37} and
\eqref{eq:38}, into \eqref{eq:40} we solve \eqref{eq:40} with
respect to $x_{2}(\tau)$, and that for an gradually increasing $k
= [1,\ldots,40]$. Note that an increasing $k$ provides solely an
increased accuracy in evaluating the matrix exponential
\eqref{eq:41}. For all the solutions evaluated with the help of
the Symbolic Math Toolbox\textsuperscript{TM}, it is found that
\eqref{eq:40} has no initial-value solution other than zero,
meaning $x_{2}(\tau) = x_{2}(t_{s,n}) = 0$. That means there are
no other initial conditions than zero for which a stick-slip cycle
could lead to $x(t) \in \Lambda$ at $t = t_{s,n+1}$. This
contradicts our initial assumption that such initial conditions
exist and, hence, completes the proof.

\end{proof}

\begin{remark}
Since no relative motion occurs during a stiction phase, cf.
Section \ref{sec:2:sub:2}, the trajectory solution \eqref{eq:40}
represents the single descriptor of the system dynamics, which is
determining convergence during the slipping phases. One can
recognize that the discontinuous Coulomb friction contributes as a
constant piecewise-continuous input $u$ to the solution of
trajectories $x(t)$ at $t_{c,n} < t < t_{s,n+1}$. Thus, it comes
as not surprising that the stick-slip convergence appears only
asymptotically, meaning either within one or a (theoretically)
infinite number of the stick-slip cycles. We also note that this
is independent of whether $x(t)$ reattains $S_0$ with or without
overshooting of $x_2=0$.
\end{remark}

\section{Numerical examples}
\label{sec:4}

The following numerical examples serve to illustrate and evaluate
the above analysis. A dedicated numerical simulation of the
stick-slip dynamics is developed by implementing \eqref{eq:21},
\eqref{eq:22} and \eqref{eq:30}, while the conditions of Theorem
\ref{thm:1} provide switching between the piecewise smooth
trajectories of the alternating slipping and sticking phases of
the relative motion of system \eqref{eq:3}-\eqref{eq:7}.

\begin{eexample}
\label{exmp:2}

Consider the system \eqref{eq:3}-\eqref{eq:7} with $K_{d}=20$,
$K_{p}=100$, $K_{i}=1000$ and varying $F_{c}=\{50, 75, 100\}$. The
initial values are assigned as $x(0)=\{0, -1.1, 0\}$,
corresponding to a classical positioning task for the
feedback-controlled system \eqref{eq:1}. Note that
$|x_{2}(0)|>F_{c}K_{p}^{-1}$ so that the trajectories start
outside of $S_{0}$ and are, therefore, inherently in the slipping
phase. The transient and convergence responses for all three
Coulomb friction values are shown opposite to each other in Figure
\ref{fig:5}, cf. qualitatively with an experimental convergence
pattern reported in \cite[Fig~4]{beerens2019}.
\begin{figure}[!h]
\centering \footnotesize (a)
\includegraphics[width=0.6\columnwidth]{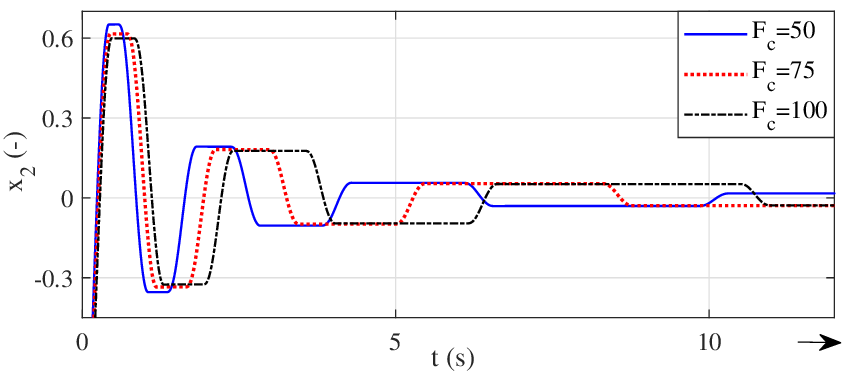}
(b)
\includegraphics[width=0.6\columnwidth]{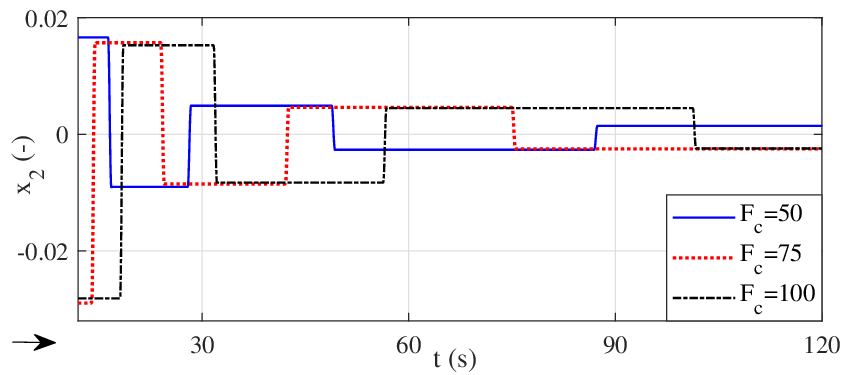}
\normalsize \caption{Output response of Example \ref{exmp:2}:
transient phase $t=[0,\ldots,12]$ sec (a), convergence phase
$t=[12,\ldots,120]$ sec (b)} \label{fig:5}
\end{figure}

\end{eexample}

\begin{eexample}
\label{exmp:3}

Consider the system \eqref{eq:3}-\eqref{eq:7} with $K_{d}=10$,
$K_{p}=1040$, $K_{i}=8000$ and $F_c=100$. The initial values
$x(0)=(0, -0.15, 0)$ are assigned to be close to, but still
outside of, the $S_{0}$-region. The linear damping $K_{d}$ is
selected with respect to $F_{c}$, so that the system exhibits only
one initial overshoot; and the stick and slip phases alternate
without changing the sign of $x_{2}$. The output displacement
response is shown in Figure \ref{fig:6}(a). The stick-slip
convergence without zero-crossing is particularly visible on the
logarithmic scale in Figure \ref{fig:6}(b). Note that after the
series of stick-slip cycles, a further evaluation of the
alternating dynamics (about $10^{-13}$ in order of magnitude) is
no longer feasible, due to a finite time step and corresponding
numerical accuracy, cf. 1st quadrant of the $S_{0}$-rhombus in
Figure \ref{fig:2}.
\begin{figure}[!h]
\centering \footnotesize (a)
\includegraphics[width=0.6\columnwidth]{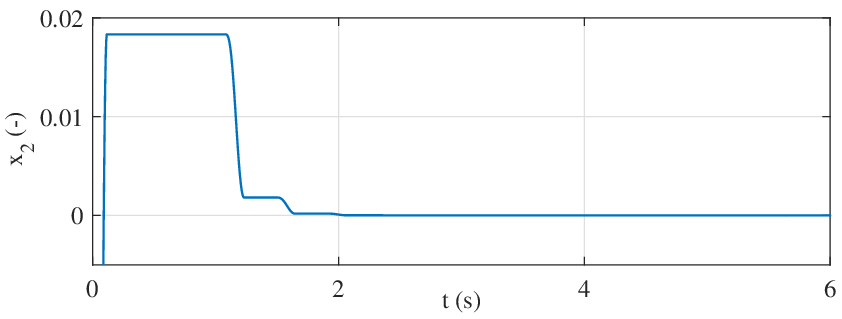}
(b)
\includegraphics[width=0.6\columnwidth]{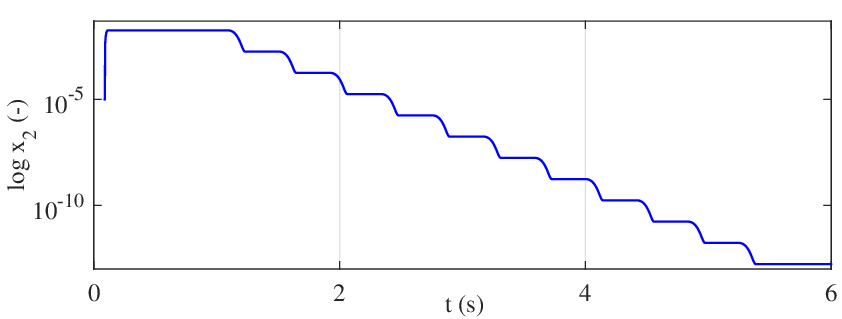}
\normalsize \caption{Output (a) and its logarithmic (b) response
of Example \ref{exmp:3}} \label{fig:6}
\end{figure}

\end{eexample}

\begin{eexample}
\label{exmp:4}

Consider the system \eqref{eq:3}-\eqref{eq:7} with $K_{d}=56$,
$K_{p}=1040$, $K_{i}=6400$ and $F_{c}=100$. Note that the control
gains are assigned in such a way that the linear dynamics
\eqref{eq:3} and \eqref{eq:6} reveal a double real pole at
$\lambda_{1,2}=-20$ and the third one in its vicinity at
$\lambda_{3}=-16$. This ensures that all states converge fairly
simultaneously towards zero, once the $\mathrm{sign}(x_{3})$
remains unchanged. For the initial conditions, $x_{1}(0),
x_{3}(0)=0$ and the varying initial displacements
$x_{2}(0)=\{-0.2, -0.25, -0.3, -0.35\}$ are assumed. Note that all
$x(0)$ are outside of $S_{0}$, while the transient overshoot lands
(in all cases) within $S_{0}$, thus directly leading to the first
stiction after an overshoot; see Figure \ref{fig:7}. One can
recognize that the integral state requires, then, the quite
different times before the system passes again into the slipping.
During the slipping phase, all states converge asymptotically
towards zero, provided $F_{c}$ remains constant. Here, it is
important to notice that in the real physical systems, a varying
$F_{c}$-value and the so-called frictional adhesion, see e.g.
\cite{zeng2006}, at extremely low velocities, will both lead to
the system passing into a sticking phase again, therefore,
provoking rather the multiple stick-slip cycles. Even though it is
not a case here with our ideal Coulomb friction assumption, the
Theorem \ref{thm:2} still holds, since there is only an asymptotic
convergence after at least one stick-slip cycle occurred.
\begin{figure}[!h]
\centering
\includegraphics[width=0.6\columnwidth]{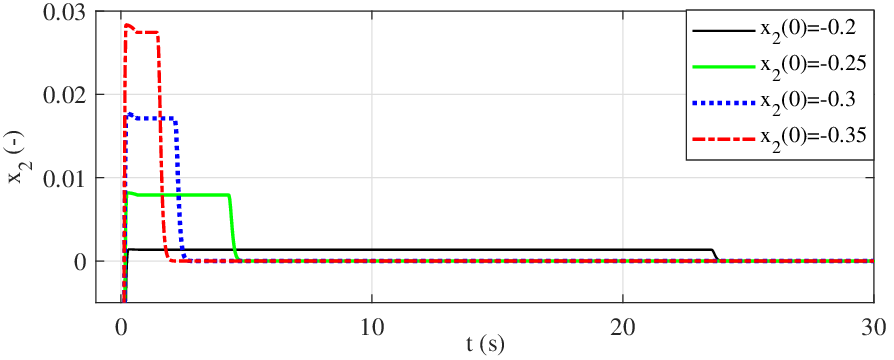}
\caption{Output response of Example \ref{exmp:4} for different
initial values} \label{fig:7}
\end{figure}

\end{eexample}

\begin{eexample}
\label{exmp:5}

Consider the system \eqref{eq:3}-\eqref{eq:7} with $K_{d}=20$,
$K_{p}=100$, $K_{i}=1000$ and $F_{c}=50$. The initial condition
$x(0)=(0, -0.5, 0)$ is assigned to be on the boundary of $S_{0}$,
thus leading to a short initial slipping and, then, providing a
large number of the stick-slip cycles by a long-term simulation
with $t=[0,\ldots,100000]$ sec. The output is shown as logarithmic
absolute value (due to the alternating sign) over the logarithmic
time argument in Figure \ref{fig:8}. One can recognize that each
consequent sticking phase proceeds closer to the origin, while the
stick-slip period grows exponentially, cf. the logarithmic
timescale. This further confirms an asymptotic convergence within
the stick-slip cycles, cf. Theorem \ref{thm:2}.
\begin{figure}[!h]
\centering
\includegraphics[width=0.6\columnwidth]{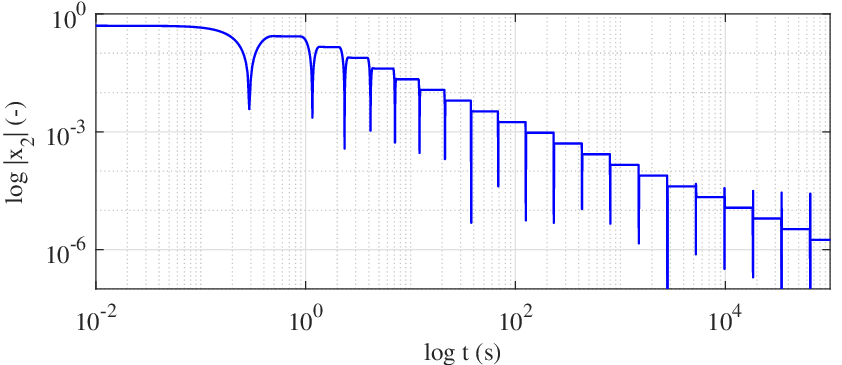}
\caption{Output response of Example \ref{exmp:5}, the logarithmic
absolute value over the logarithmic time argument} \label{fig:8}
\end{figure}

\end{eexample}

\section{Conclusions}
\label{sec:6}

An analysis of the stick-slip behavior during settling of the
feedback-controlled motion with Coulomb friction has been
developed. The most general case of a frictional discontinuity at
velocity zero-crossing has been assumed, and the parametric
conditions for appearance of a stiction region, encompassing the
equilibrium set, have been derived, that independent of the
initial conditions.

To notice is that a symmetric Coulomb friction about the origin
(i.e. zero velocity) is considered. For an asymmetric friction,
i.e. with different Coulomb friction coefficients $F_{c,\{p,n\}}$
for positive (p) and negative (n) direction, the provided analysis
is equally valid and requires solely a separate trajectories
evaluation for $x_3 > 0$ and $x_3 < 0$, cf. Section \ref{sec:3}.
The aspects of Stribeck friction (see e.g. \cite{armstrong1994}
for details) are not accounted as less relevant for principal
stick-slip behavior, even though they certainly affect the period
and shape of the corresponding stick-slip cycles. Here we recall
that the Stribeck effect provides a short-term transient negative
damping and, therefore, rather contributes to the fact that the
trajectories each time leave earlier a stiction region, before
(unavoidably) coming back into stiction at $x_2 \neq 0$.

Theorem \ref{thm:1} and Corollary \ref{clr:1} proved the stiction
region to be globally reachable and attractive. Theorem
\ref{thm:2} stated that the convergence is only asymptotically
possible and occurs with at least one but mostly an infinite
number of the stick-slip cycles in a sequence. In particular, an
'ideal' convergence of the control configuration with all real
poles in a neighborhood to each other appears with one initial
stick-slip cycle, followed by an asymptotic convergence without
new stick-slip transitions. The number of illustrative numerical
examples, with different initial conditions and parameter
settings, argue in favor of the developed analysis and provide
additional insight into the stick-slip mechanisms of a feedback
controlled motion with Coulomb friction.


\bibliographystyle{wileyNJD-AMS}
\bibliography{references}

\begin{thebibliography}{10}
\providecommand{\url}[1]{\texttt{#1}}
\providecommand{\urlprefix}{URL }
\expandafter\ifx\csname urlstyle\endcsname\relax
  \providecommand{\doi}[1]{doi:\discretionary{}{}{}#1}\else
  \providecommand{\doi}{doi:\discretionary{}{}{}\begingroup
  \urlstyle{rm}\Url}\fi

\bibitem{al2008}
F.~Al-Bender and J.~Swevers, \textit{Characterization of friction force
  dynamics}, IEEE Control Systems Magazine \textbf{28} (2008), no.~6, 64--81.

\bibitem{alvarez2000}
J.~Alvarez, I.~Orlov, and L.~Acho, \textit{An invariance principle for
  discontinuous dynamic systems with application to a {Coulomb} friction
  oscillator}, J. Dyn. Sys., Meas., Control \textbf{122} (2000), no.~4,
  687--690.

\bibitem{ang2005}
K.~H. Ang, G.~Chong, and Y.~Li, \textit{{PID} control system analysis, design,
  and technology}, IEEE transactions on control systems technology \textbf{13}
  (2005), no.~4, 559--576.

\bibitem{armstrong1996}
B.~Armstrong and B.~Amin, \textit{{PID} control in the presence of static
  friction: A comparison of algebraic and describing function analysis},
  Automatica \textbf{32} (1996), no.~5, 679--692.

\bibitem{armstrong1994}
B.~Armstrong-H{\'e}louvry, P.~Dupont, and C.~C. De~Wit, \textit{A survey of
  models, analysis tools and compensation methods for the control of machines
  with friction}, Automatica \textbf{30} (1994), no.~7, 1083--1138.

\bibitem{Awrejcewicz2005}
J.~Awrejcewicz and P.~Olejnik, \textit{{Analysis of Dynamic Systems With
  Various Friction Laws}}, ASME Applied Mechanics Reviews \textbf{58} (2005),
  no.~6, 389--411.

\bibitem{beerens2019}
R.~Beerens et~al., \textit{Reset integral control for improved settling of
  {PID}-based motion systems with friction}, Automatica \textbf{107} (2019),
  483--492.

\bibitem{bisoffi2017}
A.~Bisoffi et~al., \textit{Global asymptotic stability of a {PID} control
  system with {Coulomb} friction}, IEEE Transactions on Automatic Control
  \textbf{63} (2017), no.~8, 2654--2661.

\bibitem{canudas1987}
C.~Canudas, K.~Astrom, and K.~Braun, \textit{Adaptive friction compensation in
  dc-motor drives}, IEEE Journal on Robotics and Automation \textbf{3} (1987),
  no.~6, 681--685.

\bibitem{Clegg1958}
J.~Clegg, \textit{A nonlinear integrator for servomechanisms}, Transactions of
  the American Institute of Electrical Engineers, Part II: Applications and
  Industry \textbf{77} (1958), no.~1, 41--42.

\bibitem{DeWit1995}
C.~C. De~Wit et~al., \textit{A new model for control of systems with friction},
  IEEE Transactions on automatic control \textbf{40} (1995), no.~3, 419--425.

\bibitem{edwards1998}
C.~Edwards and S.~Spurgeon, \textit{Sliding mode control: theory and
  applications}, CRC Press, 1998.

\bibitem{filippov1988}
A.~Filippov, \textit{Differential equations with discontinuous right-hand
  sides}, 1988.

\bibitem{hensen2003}
R.~H. Hensen, M.~Van~de Molengraft, and M.~Steinbuch, \textit{Friction induced
  hunting limit cycles: A comparison between the {LuGre} and switch friction
  model}, Automatica \textbf{39} (2003), no.~12, 2131--2137.

\bibitem{johansson1999}
K.~H. Johansson, A.~Rantzer, and K.~J. {\AA}str{\"o}m, \textit{Fast switches in
  relay feedback systems}, Automatica \textbf{35} (1999), no.~4, 539--552.

\bibitem{Karnopp1985}
D.~Karnopp, \textit{Computer simulation of stick-slip friction in mechanical
  dynamic systems}, Journal of dynamic systems, measurement, and control
  \textbf{107} (1985), no.~1, 100--103.

\bibitem{khalil2002}
H.~Khalil, \textit{Nonlinear Systems}, 3rd edn., Prentice Hall, 2002.

\bibitem{koizumi1984}
T.~Koizumi and H.~Shibazaki, \textit{A study of the relationships governing
  starting rolling friction}, Wear \textbf{93} (1984), no.~3, 281--290.

\bibitem{olsson2001}
H.~Olsson and K.~J. Astrom, \textit{Friction generated limit cycles}, IEEE
  Transactions on Control Systems Technology \textbf{9} (2001), no.~4,
  629--636.

\bibitem{putra2007}
D.~Putra, H.~Nijmeijer, and N.~van~de Wouw, \textit{Analysis of
  undercompensation and overcompensation of friction in {1DOF} mechanical
  systems}, Automatica \textbf{43} (2007), no.~8, 1387--1394.

\bibitem{radcliffe1990}
C.~J. Radcliffe and S.~C. Southward, \textit{A property of stick-slip friction
  models which promotes limit cycle generation}, \textit{American Control
  Conference, 1990}, 1198--1205.

\bibitem{ruderman2017}
M.~Ruderman, \textit{On break-away forces in actuated motion systems with
  nonlinear friction}, Mechatronics \textbf{44} (2017), 1--5.

\bibitem{ruderman2015}
M.~Ruderman and M.~Iwasaki, \textit{Observer of nonlinear friction dynamics for
  motion control}, IEEE Transactions on Industrial Electronics \textbf{62}
  (2015), no.~9, 5941--5949.

\bibitem{ruderman2016}
M.~Ruderman and M.~Iwasaki, \textit{Analysis of linear feedback position
  control in presence of presliding friction}, IEEJ Journal of Industry
  Applications \textbf{5} (2016), no.~2, 61--68.

\bibitem{ruderman2020}
M.~Ruderman, M.~Iwasaki, and W.-H. Chen, \textit{Motion-control techniques of
  today and tomorrow: A review and discussion of the challenges of controlled
  motion}, IEEE Industrial Electronics Magazine \textbf{14} (2020), no.~1,
  41--55.

\bibitem{shtessel2014}
Y.~Shtessel et~al., \textit{Sliding mode control and observation}, Springer,
  2014.

\bibitem{symens2005}
W.~Symens and F.~Al-Bender, \textit{Dynamic characterization of hysteresis
  elements in mechanical systems. {II.} experimental validation}, Chaos: An
  Interdisciplinary Journal of Nonlinear Science \textbf{15} (2005), no.~1,
  013106.

\bibitem{yoon2019}
J.~Y. Yoon and D.~L. Trumper, \textit{Friction microdynamics in the time and
  frequency domains: Tutorial on frictional hysteresis and resonance in
  precision motion systems}, Precision Engineering \textbf{55} (2019),
  101--109.

\bibitem{zames1966}
G.~Zames, \textit{On the input-output stability of time-varying nonlinear
  feedback systems part one: Conditions derived using concepts of loop gain,
  conicity, and positivity}, IEEE transactions on automatic control \textbf{11}
  (1966), no.~2, 228--238.

\bibitem{zeng2006}
H.~Zeng, M.~Tirrell, and J.~Israelachvili, \textit{Limit cycles in dynamic
  adhesion and friction processes: a discussion}, The Journal of Adhesion
  \textbf{82} (2006), no.~9, 933--943.

\end{thebibliography}

\section*{Author Biography}

\begin{biography}{}{\textbf{Michael Ruderman}
earned his Dr.-Ing. degree in electrical engineering from TU
University Dortmund, Germany, in 2012. He is a full professor at
the University of Agder, Grimstad, Norway, teaching control theory
in B.Sc., M.Sc., and Ph.D. degree programs. He serves in different
editorial boards and technical committees of IEEE and IFAC
societies and is chairing IEEE/IES TC on Motion Control in the
terms 2018-2019 and 2020-2021. He is a Senior Member of IEEE and
was the general chair of the 16th IEEE International Workshop on
Advanced Motion Control, in 2020. His current research interests
are in motion control, nonlinear dynamics, and hybrid control
systems.}
\end{biography}

\end{document}